\documentclass[11pt, a4paper]{article}
\usepackage[a4paper, top=2.5cm, bottom=2.5cm, left=2cm, right=2cm]{geometry}
\usepackage{amsmath, amssymb, amsthm}
\usepackage{graphicx}
\usepackage{hyperref}
\usepackage{cite}
\usepackage{placeins}
\usepackage{booktabs}

\hypersetup{
    colorlinks=true,
    linkcolor=blue,
    citecolor=blue,
    urlcolor=blue
}

\title{\textbf{Lobe Dynamics, Phase-Space Transport, and Non-Adiabatic Leakage Thresholds in the Nonautonomous Kerr-Cat Qubit}}
\author{Stephen Wiggins \\ 
\small Hetao Institute of Mathematics and Interdisciplinary Sciences, Shenzhen, China \\
\small School of Mathematics, University of Bristol, UK}
\date{\today}

\begin{document}

\maketitle

\begin{abstract}
The Kerr-nonlinear parametric oscillator (KPO) provides a foundational semiclassical model
for cat-state quantum hardware. Standard analyses of the KPO typically rely on autonomous,
frozen-time approximations to describe the stabilization of macroscopic coherent states. However,
state preparation and gate manipulation are driven by explicitly time-dependent microwave pulses,
so the operational dynamics are inherently nonautonomous. In this paper, we show that static
algebraic equilibrium pictures are incomplete for describing both state formation and gate-induced
transport in the Kerr-cat qubit. For nonautonomous state preparation, we analyze the ramped
resonant model by combining a linear nonautonomous stability analysis with a local invariant-graph
reduction near the vacuum trajectory. This yields a quintic reduced normal form in the critical
direction and identifies two symmetric post-threshold moving branches that organize the local
state-formation dynamics. The associated diagnostics separate the reduced branch dynamics from
the full two-dimensional phase-twist relaxation observed in the hardware coordinates. For gate
execution, we model a fast pulse as a weak aperiodic perturbation of the conservative resonant
figure-eight separatrix and apply Melnikov's method to derive a leading-order transport criterion.
In this framework, transient lobe dynamics emerge as a semiclassical mechanism for non-adiabatic
leakage, and the resulting amplitude-width threshold curve provides a leading-order geometric
indicator for the onset of gate-pulse-induced transport.

\vspace{1em}
\end{abstract}

\section{Introduction}

The realization of hardware-efficient quantum error correction has generated intense interest in
continuous-variable quantum systems, with the Kerr-cat qubit emerging as a leading architecture
\cite{Mirrahimi2014,Goto2016}. By driving a Kerr-nonlinear oscillator with a two-photon
parametric pump, the phase space of the system can be engineered to exhibit two degenerate,
macroscopic coherent states. These states, stabilized by continuous dissipation, serve as the logical
computational basis, offering exponential suppression of bit-flip errors.

Despite the elegance of this architecture, executing fast, high-fidelity quantum gates remains a
significant experimental challenge. Physical gate operations and state initialization are driven by
explicitly time-dependent electromagnetic microwave pulses. Much of the existing quantum optics
and hardware literature relies on autonomous, frozen-time or adiabatic pictures when discussing
state stability. A common approach is to analyze an effective static potential or frozen-time
equilibria, sometimes supplemented by Fokker--Planck or activation-type descriptions
\cite{ConsaniWarburton2020,Dykman2012}.

Time-dependence is generally treated only via the quantum adiabatic theorem, assuming the system
tracks an instantaneous sequence of static ground states. This autonomous intuition is useful but
insufficient for operational control, and can become misleading if frozen-time equilibria are
interpreted as physical trajectories. In an explicitly time-dependent, driven-dissipative system, the
instantaneous frozen-time zeros of the vector field can provide useful reference curves, but they do
not themselves satisfy the equations of motion and therefore do not describe the evolving phase-space
objects that organize transport. As demonstrated in classical transition state theory, static barrier
models can fail to capture the actively moving phase-space bottlenecks that govern transport under
fast external driving \cite{Wiggins2025,Malhotra1998}.

In this paper, we bridge this methodological gap by importing the geometric framework of
nonautonomous dynamical systems into the operational physics of the Kerr-cat qubit. While
nonautonomous bifurcations, invariant sets, and transient transport have been studied extensively
in the applied mathematics literature \cite{Kloeden2011,Langa2002,Rasmussen2007},
these tools are rarely deployed in the context of superconducting quantum hardware design.

The paper follows a two-stage modeling strategy. Section~2 formulates the nonautonomous
semiclassical model and reviews the standard stationary Kerr-cat picture. Section~3 constructs the
conservative resonant figure-eight skeleton that later supports the Melnikov analysis. Section~4
revisits state preparation under a nonautonomous ramp: the linearized dynamics identify the
critical direction, and a local invariant-graph reduction yields a quintic reduced normal form
together with two symmetric post-threshold moving branches of the reduced dynamics. The
preparation diagnostics then separate the reduced scalar branch dynamics from the full
$x$--$y$ hardware trajectories, clarifying the role of Kerr-induced transverse phase twist. Section~5
turns to gate execution, modeling a fast pulse as a weak aperiodic perturbation of the resonant
separatrix and deriving a leading-order Melnikov threshold for the onset of transient
lobe-mediated transport. Section~6 summarizes the scope of the results and outlines directions
for future work.

\section{The Nonautonomous Kerr-Cat Model}

To ground the mathematical analysis in physical reality, we consider the foundational architecture
of the Kerr-cat qubit: a superconducting nonlinear resonator subjected to a two-photon parametric
drive~\cite{Mirrahimi2014}.

\subsection{Derivation of the Model and Hardware Parameters}

The fundamental dynamics are governed by a driven-dissipative open quantum system. In the large
photon-number limit, we can employ a formal semiclassical expansion where the quantum annihilation
operator $\hat{a}$ is replaced by a continuous complex amplitude $\alpha\in\mathbb{C}$~\cite{Dykman2012}.
We analyze the system in a frame rotating at half the parametric pump frequency
($\omega_d=\omega_p/2$). Applying the Rotating Wave Approximation (RWA) to average out
fast-oscillating terms~\cite{ConsaniWarburton2020}, the evolution of the intra-cavity field is governed by the
following nonautonomous differential equation:
\begin{equation}\label{eq:complex_kpo}
\dot{\alpha}
=
-\left(\frac{\kappa}{2}+i\Delta\right)\alpha
-iK|\alpha|^2\alpha
+p(t)\alpha^\ast .
\end{equation}
The phase space geometry is dictated by the interplay of four physical parameters:
\begin{itemize}
    \item $\kappa>0$ is the single-photon dissipation rate, representing engineered and environmental
    photon loss from the cavity. It contracts the global phase space volume, rendering the system
    non-conservative.

    \item $\Delta=\omega_a-\omega_d$ is the detuning between the bare cavity resonance $\omega_a$ and
    the rotating-frame frequency.

    \item $K>0$ is the Kerr nonlinearity, originating from the anharmonicity of the Josephson
    junctions. This term is responsible for amplitude-dependent frequency shifts and provides
    nonlinear confinement.

    \item $p(t)$ is the parametric pump envelope. In the hardware, this is an explicitly
    time-dependent microwave voltage control field. It acts as a two-photon drive that forces energy
    into the system. For all physical initialization protocols discussed in this paper, we assume the
    envelope is non-negative, $p(t)\ge 0$.
\end{itemize}

Decomposing the complex amplitude into Cartesian real and imaginary coordinates,
\[
\alpha=x+iy,
\]
yields the planar nonautonomous vector field
\begin{equation}\label{eq:x_dot}
\dot{x}
=
\left(p(t)-\frac{\kappa}{2}\right)x
+\Delta y
+Ky(x^2+y^2),
\end{equation}
\begin{equation}\label{eq:y_dot}
\dot{y}
=
-\left(p(t)+\frac{\kappa}{2}\right)y
-\Delta x
-Kx(x^2+y^2).
\end{equation}

\subsection{The Autonomous Baseline: Stationary Bifurcations}

Before addressing explicit time dependence, it is useful to recall the standard stationary Kerr-cat
picture obtained by freezing the pump at a constant value $p(t)=p_0$. Equations~\eqref{eq:x_dot}
and~\eqref{eq:y_dot} then define an autonomous planar flow.

The vacuum state, represented by the origin $(0,0)$, is a fixed point for all $p_0$. Its local stability
is determined by the Jacobian at the origin,
\begin{equation}
J_0
=
\begin{pmatrix}
p_0-\dfrac{\kappa}{2} & \Delta \\[4pt]
-\Delta & -\left(p_0+\dfrac{\kappa}{2}\right)
\end{pmatrix}.
\end{equation}
The eigenvalues are
\[
\lambda_\pm
=
-\frac{\kappa}{2}
\pm
\sqrt{p_0^2-\Delta^2}.
\]
Accordingly, the linearized threshold occurs at
\[
p_c
=
\sqrt{\Delta^2+(\kappa/2)^2}.
\]
For $p_0<p_c$, the origin is locally asymptotically stable. For $p_0>p_c$, the origin is of saddle type.

Within the standard stationary Kerr-cat bifurcation picture~\cite{Goto2016,Grimm2020}, this local
threshold is associated with the appearance of two additional symmetry-related stable states. In the
resonant case these are aligned with the natural coherent-state directions, while for nonzero detuning
their phases are shifted. These nontrivial fixed points correspond to the macroscopic coherent states
$|+\alpha\rangle$ and $|-\alpha\rangle$ that form the logical basis of the device. Thus the frozen-time
autonomous system provides the usual intuition of two separated logical wells, with the origin acting
as the dividing saddle once the pump exceeds threshold.

In the quantum optics literature, a coherent state is the closest quantum mechanical analogue to a
classical oscillating electromagnetic field, with the complex parameter $\alpha$ determining its
macroscopic amplitude and phase. Because the underlying hardware is a continuous-variable resonator
with an infinite-dimensional Hilbert space, operation as a qubit requires dynamically isolating an
effective two-dimensional subspace. In the stationary Kerr-cat picture, the two symmetry-related
attracting equilibria are mapped to the computational basis states
\[
|0\rangle\equiv |+\alpha\rangle,
\qquad
|1\rangle\equiv |-\alpha\rangle.
\]

The architecture is called a ``Kerr-cat'' because true quantum superpositions of these two
macroscopically distinct coherent fields can also be formed. In practice, these are the even and odd
cat states,
\[
|C_\pm\rangle \propto |+\alpha\rangle \pm |-\alpha\rangle.
\]
At the semiclassical level considered here, however, the main point is simpler: the frozen-time
autonomous model suggests a pair of separated logical regions, but says little about how those states
are dynamically created under a ramped control pulse. This distinction is essential for the remainder
of the paper. The autonomous picture identifies the operating objective of the hardware, but the
actual preparation protocol ramps $p(t)$ from $0$ to $p_{\max}$. Since $p(t)$ changes the vector field
itself, the preparation dynamics must be analyzed as a nonautonomous two-parameter process
$\varphi_{t,t_0}(x_0,y_0)$ rather than as a sequence of exact equilibria.

\section{The Hamiltonian Limit: Pitchfork Bifurcations and Homoclinic Geometry}

Realistically, the operational qubit involves small dissipation and small time-dependent driving. To
analyze the global dynamics analytically as a function of these parameters via the Melnikov method,
we first establish the strictly conservative, autonomous limit as our geometric skeleton. We eliminate
environmental coupling by setting the single-photon dissipation to zero, $\kappa=0$, and fix the
parametric pump to a constant amplitude, $p(t)=p_0$.

\subsection{The Effective Hamiltonian and Pitchfork Bifurcation}

To generate the undamped limit of the vector field defined in Equations~\eqref{eq:x_dot}
and~\eqref{eq:y_dot}, the rotating-frame dynamics are governed by the effective Hamiltonian
\begin{equation}\label{eq:effective_hamiltonian}
H(x,y)
=
\frac{\Delta}{2}(x^2+y^2)
+
\frac{K}{4}(x^2+y^2)^2
+
p_0xy .
\end{equation}
The Hamiltonian equations of motion,
\[
\dot{x}=\frac{\partial H}{\partial y},
\qquad
\dot{y}=-\frac{\partial H}{\partial x},
\]
yield the unperturbed KPO phase space, where the topology depends on the parametric drive strength
$p_0$ relative to the detuning $\Delta$.

The origin $(0,0)$ remains a fixed point for all parameter values, representing the vacuum state.
Because the energy at the origin is $H(0,0)=0$, its stability is dictated by the eigenvalues of the
linearized Hamiltonian vector field,
\[
\lambda_\pm=\pm\sqrt{p_0^2-\Delta^2}.
\]
When $p_0<|\Delta|$, the eigenvalues are purely imaginary, and the vacuum state is a neutrally
stable elliptic center.

However, when the drive exceeds the critical threshold $p_0>|\Delta|$, the origin undergoes a
Hamiltonian pitchfork bifurcation~\cite{Wiggins2003}. The eigenvalues become real, transforming the
origin into a hyperbolic saddle point. Simultaneously, two new degenerate elliptic centers are born.
These centers represent the unperturbed macroscopic coherent states, $|+\alpha\rangle$ and
$|-\alpha\rangle$. In the operational device, they correspond to classical-like continuous-wave
microwave fields with opposite phases, establishing the logical $|0\rangle$ and $|1\rangle$
computational basis. The architecture's namesake ``cat states'' are the true quantum superpositions
of these two classical limit points, a phenomenon requiring a full quantum Hilbert-space treatment
beyond this semiclassical expansion.

\subsection{The Bounded Lemniscate and Figure-Eight Topology}

To determine the global manifold structure associated with the saddle at the origin, we rely on the
boundedness of the critical energy contour. In the parametrically resonant case, $\Delta=0$, the
effective Hamiltonian in the symplectic, rotated coordinates
\[
(X,Y)
=
\frac{1}{\sqrt{2}}(x+y,-x+y)
\]
decouples the quadratic terms:
\begin{equation}\label{eq:rotated_hamiltonian}
H(X,Y)
=
\frac{K}{4}(X^2+Y^2)^2
+
\frac{p_0}{2}(X^2-Y^2).
\end{equation}

\begin{figure}[!htbp]
    \centering
    \includegraphics[width=0.75\textwidth]{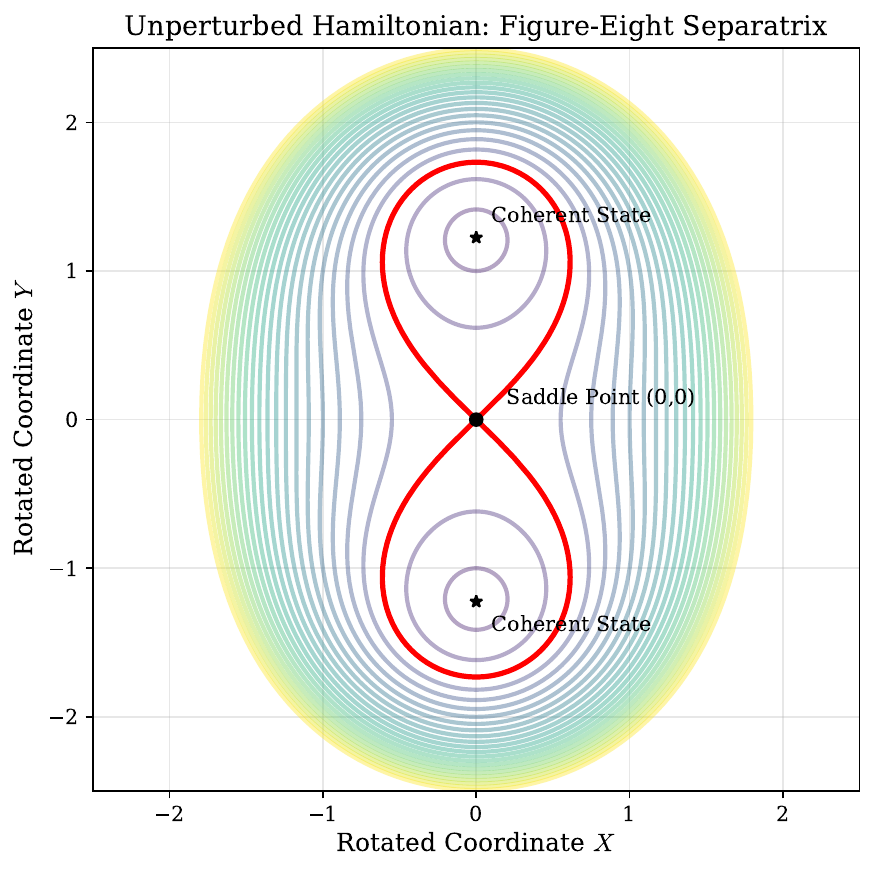}
    \caption{\small Unperturbed conservative skeleton in the rotated coordinate frame
    $(X,Y)$. The red $H=0$ contour is the bounded figure-eight separatrix of the resonant
    Hamiltonian system. The saddle at the origin is marked explicitly, and the surrounding
    Hamiltonian contours show the two lobe regions associated with the two macroscopic coherent-state
    wells. Because the $H=0$ contour is bounded, the unstable branches emerging from the saddle are
    forced to return to the saddle and coincide with the stable branches, forming the two homoclinic
    loops used as the unperturbed skeleton for the Melnikov calculation. Representative parameters
    are $K=1.0$ and $p_0=1.5$.}
    \label{fig:unperturbed_separatrix}
\end{figure}

Because the system is conservative, $\kappa=0$, the stable and unstable manifolds associated with
the saddle point at the origin must lie entirely on the $H(0,0)=0$ energy contour. Furthermore, in
this rotated frame, setting the vector field to zero forces $X=0$, isolating the macroscopic coherent
states on the vertical axis at
\[
(0,\pm\sqrt{p_0/K}).
\]
To demonstrate that the invariant manifolds do not escape to infinity, we convert the Hamiltonian
into polar coordinates $(R,\theta)$, where
\[
X=R\cos\theta,
\qquad
Y=R\sin\theta.
\]
Substituting these into the Hamiltonian and setting $H=0$ yields the geometric equation for the
saddle's level set:
\begin{equation}
R^2
\left(
\frac{K}{4}R^2
+
\frac{p_0}{2}\cos(2\theta)
\right)
=
0.
\end{equation}
Discarding the trivial $R=0$ point at the origin, the equation for the nontrivial contour is
\begin{equation}
R^2
=
-\frac{2p_0}{K}\cos(2\theta).
\end{equation}
Because $R^2$ must be non-negative, the contour only exists in the angular sectors where
$\cos(2\theta)\le 0$, specifically
\[
\theta\in[\pi/4,3\pi/4]
\qquad\text{and}\qquad
\theta\in[5\pi/4,7\pi/4].
\]
As illustrated in Figure~\ref{fig:unperturbed_separatrix}, this defines a bounded lemniscate---a
figure-eight curve with maximum radius
\[
R_{\max}=\sqrt{\frac{2p_0}{K}}.
\]

Because the unstable branches emerging from the saddle are confined to this bounded,
one-dimensional closed contour, they cannot escape to infinity. They are geometrically forced to
return to the saddle as $t\to\infty$, coinciding with the stable manifolds to form a pair of
unperturbed homoclinic orbits.

For the subsequent Melnikov analysis, we rely on the topological properties of these unperturbed
homoclinic trajectories evaluated in the original $(x,y)$ hardware frame. Due to the integrability
and reflectional symmetries of the Hamiltonian, the right lobe of the figure-eight can be
parameterized by $(x_h(t),y_h(t))$ such that $x_h(t)$ is an even function of time,
\[
x_h(-t)=x_h(t),
\]
and $y_h(t)$ is an odd function of time,
\[
y_h(-t)=-y_h(t).
\]

The geometric phase-space area enclosed by this lobe is positive. For the oriented homoclinic
parameterization used in the Melnikov calculation below, we write
\begin{equation}\label{eq:oriented_lobe_area}
\frac{1}{2}
\int_{-\infty}^{\infty}
\bigl(x_h\dot{y}_h-y_h\dot{x}_h\bigr)\,dt
=
-A_{\mathrm{lobe}},
\qquad
A_{\mathrm{lobe}}>0.
\end{equation}
Thus $A_{\mathrm{lobe}}$ denotes the positive geometric area of the lobe, while the signed integral
appearing in the Melnikov calculation carries the sign determined by the chosen orientation of the
homoclinic orbit. Reversing the homoclinic parameterization reverses this signed integral and
multiplies the Melnikov function by an overall sign, but does not change the zero condition for
transversal intersections.

We emphasize that all subsequent preparation and transport calculations in Sections~4 and~5 are
interpreted and evaluated in the original physical hardware coordinates $(x,y)$. The rotated frame
$(X,Y)$ is introduced here solely to expose the underlying topology of the conservative skeleton.

\FloatBarrier

\section{Local Invariant-Graph Reduction and Moving Branches}

The initialization of the Kerr-cat qubit requires ramping the parametric pump envelope $p(t)$ from
zero to a maximum operational value $p_{\max}$. In the applied quantum hardware literature, this
process is frequently analyzed by treating $p(t)$ as a constant parameter at each instant in time and
setting the vector field to zero to find the instantaneous frozen-time equilibria.

However, in an explicitly time-dependent system, this algebraic root-finding fails to capture the true
dynamics. While one can always algebraically solve for the time-dependent roots of the vector field
components, these resulting curves do not satisfy the differential equations and are not invariant
trajectories. To properly model how the macroscopic logical states physically emerge from the
0-photon vacuum state via a nonautonomous pitchfork bifurcation, we must utilize the geometric
tools of nonautonomous dynamical systems.

To make the analytical reductions mathematically tractable, we restrict our focus to the
parametrically resonant case $\Delta=0$. This isolates the critical mechanism in its simplest form. For
sufficiently small nonzero detuning, the hyperbolic regimes away from threshold persist by the
roughness of exponential dichotomies; however, the transition itself is most transparently analyzed in
the resonant case~\cite{Coppel1978}.

\subsection{The Variational Equation and Failure of a Global Dichotomy}

To identify the linear threshold associated with the birth of the logical states, we analyze the
stability of the vacuum trajectory. In the resonant case $\Delta=0$, the vacuum state, represented by
the origin $(x,y)=(0,0)$, remains an exact solution of the full nonautonomous system for every time:
\[
(\hat{x}(t),\hat{y}(t))=(0,0).
\]
For an autonomous system, hyperbolicity of an equilibrium is characterized by the absence of spectrum
on the imaginary axis. For a nonautonomous trajectory, the standard replacement is the existence of
an exponential dichotomy for the variational equation, in the sense of Coppel~\cite{Coppel1978}. If
\[
\dot{z}=A(t)z
\]
has evolution operator $\Phi(t,s)$, then an exponential dichotomy on $\mathbb{R}$ means that there
exist a projection $P$, constants $K\ge 1$ and $\alpha>0$, such that
\[
\|\Phi(t,s)P\|\le K e^{-\alpha(t-s)}
\qquad \text{for } t\ge s,
\]
and
\[
\|\Phi(t,s)(I-P)\|\le K e^{-\alpha(s-t)}
\qquad \text{for } t\le s.
\]
Geometrically, this means that the linearized dynamics admit a uniform stable/unstable splitting on
the whole time axis, with fixed subspace dimensions and uniform exponential estimates. This is a
global property of the linearized dynamics on the whole time axis $\mathbb{R}$; it is not a frozen-time
statement about the sign of an instantaneous coefficient.

Defining the local displacement coordinates $(\xi,\eta)$, the variational equation evaluated at the
origin is
\begin{equation}
\begin{pmatrix}
\dot{\xi}\\
\dot{\eta}
\end{pmatrix}
=
\begin{pmatrix}
p(t)-\dfrac{\kappa}{2} & 0\\[4pt]
0 & -p(t)-\dfrac{\kappa}{2}
\end{pmatrix}
\begin{pmatrix}
\xi\\
\eta
\end{pmatrix}.
\end{equation}
Because the linearization is diagonal, the fundamental solution matrix is also diagonal, and the two
coordinates evolve independently:
\begin{equation}
\xi(t)=\xi_0
\exp\left(\int_{t_0}^{t}\left(p(s)-\frac{\kappa}{2}\right)\,ds\right),
\end{equation}
\begin{equation}
\eta(t)=\eta_0
\exp\left(-\int_{t_0}^{t}\left(p(s)+\frac{\kappa}{2}\right)\,ds\right).
\end{equation}
The $\eta$-direction is straightforward. Because the physical microwave pump envelope is
non-negative, $p(t)\ge 0$, we have
\[
-\bigl(p(t)+\kappa/2\bigr)\le -\kappa/2<0
\]
for all $t$. Hence the $\eta$-component decays uniformly exponentially forward in time, so this
direction is always linearly stable.

Thus the only nontrivial issue is the $\xi$-direction. In the autonomous case where $p(t)=p_0$ is
constant, Equation~(10) reduces to
\[
\xi(t)=\xi_0 e^{(p_0-\kappa/2)(t-t_0)}.
\]
Accordingly, the origin is linearly asymptotically stable when $p_0<\kappa/2$, of saddle type when
$p_0>\kappa/2$, and nonhyperbolic at the threshold $p_0=\kappa/2$.

To model a smooth monotone turn-on while keeping the analysis explicit, we use the representative
logistic envelope:
\begin{equation}
p(t)=\frac{p_{\max}}{1+e^{-\gamma(t-t_c)}},
\end{equation}
where $p_{\max}>0$ is the maximum asymptotic pump amplitude, $\gamma>0$ defines the ramp rate,
and $t_c\in\mathbb{R}$ is the temporal center of the pulse. Substituting Equation~(12) into the
exponent of Equation~(10), we can evaluate the integral exactly:
\begin{equation}
\int_{t_0}^{t}\left(p(s)-\frac{\kappa}{2}\right)\,ds
=
\left(p_{\max}-\frac{\kappa}{2}\right)(t-t_0)
+
\frac{p_{\max}}{\gamma}
\ln\left(
\frac{1+e^{-\gamma(t-t_c)}}{1+e^{-\gamma(t_0-t_c)}}
\right).
\end{equation}
Substituting this back yields the closed-form expression for the displacement in the critical direction:
\begin{equation}
\xi(t)
=
\xi_0 e^{(p_{\max}-\kappa/2)(t-t_0)}
\left(
\frac{1+e^{-\gamma(t-t_c)}}{1+e^{-\gamma(t_0-t_c)}}
\right)^{p_{\max}/\gamma}.
\end{equation}

The explicit formula~(14) shows that the key issue is not a local ``loss of hyperbolicity at an
instant,'' but rather the incompatibility of the past and future asymptotic linear behavior. Indeed,
\[
p(t)\to 0 \quad \text{as } t\to -\infty,
\qquad
p(t)\to p_{\max} \quad \text{as } t\to +\infty.
\]
Therefore the $\xi$-direction behaves asymptotically like
\[
\dot{\xi}=-\frac{\kappa}{2}\xi
\quad \text{in the far past},
\]
and
\[
\dot{\xi}=
\left(p_{\max}-\frac{\kappa}{2}\right)\xi
\quad \text{in the far future}.
\]

This is the subtle point in the present problem. The standard Coppel definition requires a single
projection $P$ and a single stable/unstable splitting on the whole line $\mathbb{R}$. In the ramped
threshold-crossing regime, however, the critical $\xi$-direction changes its asymptotic linear type: it
is attracting in the remote past and repelling in the remote future. Thus the same direction cannot
belong to one fixed stable or unstable bundle on all of $\mathbb{R}$. The difficulty is therefore not that
the variational equation cannot be solved; rather, it is that the solution does not fit the full-line
uniformity required by the standard exponential dichotomy definition.

Two regimes follow.

\medskip
\noindent
\textbf{(i) Uniformly subcritical ramp: $p_{\max}<\kappa/2$.}
In this case,
\[
p(t)-\kappa/2 \le -\delta<0
\]
for some $\delta>0$ and all $t$. Hence both coordinates decay uniformly exponentially forward in time,
so the origin is uniformly exponentially stable. Equivalently, the variational equation admits an
exponential dichotomy on $\mathbb{R}$ with trivial unstable subspace.

\medskip
\noindent
\textbf{(ii) Ramp through threshold: $p_{\max}>\kappa/2$.}
In this operational regime, the coefficient $p(t)-\kappa/2$ is negative in the remote past and positive
in the remote future. Thus the same $\xi$-direction is attracting as $t\to -\infty$ and repelling as
$t\to +\infty$. Consequently, there is no global exponential dichotomy on $\mathbb{R}$ with a fixed-rank
splitting at the origin: the linearized dynamics do not possess a uniform hyperbolic decomposition
valid for all time.

Nevertheless, the explicit variational solution still provides a clear criterion for the relevant
change in the linear type of the vacuum trajectory in the present nonautonomous setting. The
$\eta$-direction remains uniformly stable throughout, while the $\xi$-direction changes from
asymptotically stable in the remote past to asymptotically unstable in the remote future. In this
sense, the threshold-crossing family passes from a uniformly stable full-line variational problem to
one that no longer admits a full-line exponential dichotomy with fixed rank. This is the linear
signature that signals the onset of the nonautonomous pitchfork mechanism.

Equivalently stated, on sufficiently early time intervals the linearized vacuum dynamics are of
stable type, whereas on sufficiently late time intervals they are of saddle type. Thus, although the
threshold-crossing ramp does not satisfy the full-line exponential dichotomy definition of Coppel, the
change in asymptotic linear type still identifies the relevant breakdown of uniform hyperbolicity and
isolates the critical direction, namely the $x$-axis, while the $y$-axis remains strongly stable.

However, the variational equation alone does not determine the nonlinear nature of the bifurcation.
It identifies the failure of global hyperbolicity and the one-dimensional critical direction; whether the
nonlinear transition is a pitchfork, a saddle-node, or another bifurcation is determined by the nonlinear
terms. This is the role of the invariant-graph reduction carried out next.

\subsection{Local Invariant-Graph Reduction}

We now turn from the linear analysis to the leading nonlinear feedback responsible for saturating the
growth in the critical direction. In view of the discussion in Section~4.1, the relevant reduction is a
local forward reduction near the vacuum trajectory, rather than a global statement on the whole time
axis. The role of this section is to identify the leading nonlinear term induced in the critical dynamics
by the strongly damped transverse direction once the system has entered the post-threshold regime.

We therefore restrict to the parametrically resonant case $\Delta=0$ and seek a local time-dependent
invariant graph of the form
\[
y=h(x,t),
\]
valid for trajectories that remain sufficiently close to the origin during the preparation ramp. Because
the vector field is odd under the symmetry $(x,y)\mapsto(-x,-y)$, the graph must also be odd in $x$.
The leading nontrivial term therefore has the form
\begin{equation}
y=h(x,t)=a_3(t)x^3+O(x^5).
\end{equation}
The graph must satisfy the standard nonautonomous graph-invariance equation
\begin{equation}
\dot{y}=\frac{\partial h}{\partial t}+\frac{\partial h}{\partial x}\dot{x}.
\end{equation}
Substituting the full nonlinear vector field~(2)--(3) with $\Delta=0$ and using
$y=a_3(t)x^3+O(x^5)$, the left-hand side becomes
\begin{align}
\dot{y}
&=
-\left(p(t)+\frac{\kappa}{2}\right)y-Kx(x^2+y^2) \notag\\
&=
-\left(p(t)+\frac{\kappa}{2}\right)a_3(t)x^3-Kx^3+O(x^5).
\end{align}
The right-hand side is
\begin{align}
\frac{\partial h}{\partial t}+\frac{\partial h}{\partial x}\dot{x}
&=
\dot{a}_3(t)x^3
+
3a_3(t)x^2
\left[
\left(p(t)-\frac{\kappa}{2}\right)x+Ky(x^2+y^2)
\right] \notag\\
&=
\dot{a}_3(t)x^3
+
3\left(p(t)-\frac{\kappa}{2}\right)a_3(t)x^3
+
O(x^5).
\end{align}
Matching the coefficients of $x^3$ yields
\begin{equation}
-\left(p(t)+\frac{\kappa}{2}\right)a_3(t)-K
=
\dot{a}_3(t)
+
3\left(p(t)-\frac{\kappa}{2}\right)a_3(t),
\end{equation}
or equivalently,
\begin{equation}
\dot{a}_3+(4p(t)-\kappa)a_3=-K.
\end{equation}

This equation does not by itself select a unique coefficient $a_3(t)$ on the whole time axis. Rather,
different finite choices of $a_3$ at a given time correspond to different local invariant graphs. The
appropriate way to use this equation in the present problem is therefore forward in time, starting once
the ramp has entered the post-threshold regime.

Accordingly, fix any time $T$ after the threshold crossing such that the local invariant-graph
expansion remains valid on the forward interval of interest. This time $T$ is not a distinguished
physical parameter; it simply marks the beginning of a sufficiently late post-threshold interval on
which both
\[
\mu(t)=p(t)-\kappa/2
\]
and the coefficient $4p(t)-\kappa$ are uniformly positive. Different finite choices of $T$ and of $a_3(T)$
affect only transient details of the local graph, and we show below that this dependence is forgotten
exponentially fast.

Define
\[
q(t):=4p(t)-\kappa.
\]
Because the logistic ramp is monotone and $p_{\max}>\kappa/2$ in the operational regime, we may
choose $T$ so that
\[
q(t)=4p(t)-\kappa \ge q(T)=:\delta>0
\qquad \text{for all } t\ge T.
\]
Thus, on the post-threshold interval $[T,\infty)$, the coefficient equation is a scalar linear
nonautonomous equation with a uniformly positive coefficient.

Introduce
\[
Q(t,s):=\int_s^t q(u)\,du
=
\int_s^t \bigl(4p(u)-\kappa\bigr)\,du.
\]
Then variation of constants gives the exact forward solution
\begin{equation}
a_3(t)
=
e^{-Q(t,T)}a_3(T)
-
K\int_T^t e^{-Q(t,\tau)}\,d\tau,
\qquad t\ge T.
\end{equation}
For the logistic ramp
\[
p(t)=\frac{p_{\max}}{1+e^{-\gamma(t-t_c)}},
\]
the exponent can be evaluated explicitly:
\begin{equation}
Q(t,s)
=
(4p_{\max}-\kappa)(t-s)
+
\frac{4p_{\max}}{\gamma}
\ln\left(
\frac{1+e^{-\gamma(t-t_c)}}{1+e^{-\gamma(s-t_c)}}
\right).
\end{equation}
Hence
\begin{equation}
e^{-Q(t,s)}
=
e^{-(4p_{\max}-\kappa)(t-s)}
\left(
\frac{1+e^{-\gamma(s-t_c)}}{1+e^{-\gamma(t-t_c)}}
\right)^{4p_{\max}/\gamma},
\end{equation}
and therefore
\begin{equation}
a_3(t)
=
e^{-Q(t,T)}a_3(T)
-
K\int_T^t
e^{-(4p_{\max}-\kappa)(t-\tau)}
\left(
\frac{1+e^{-\gamma(\tau-t_c)}}{1+e^{-\gamma(t-t_c)}}
\right)^{4p_{\max}/\gamma}
\,d\tau.
\end{equation}

This explicit representation immediately yields the key properties needed for the reduction. First,
the dependence on the initial graph is exponentially forgotten. If $a_3^{(1)}$ and $a_3^{(2)}$ are two
solutions corresponding to two different finite initial values at time $T$, then
\[
a_3^{(1)}(t)-a_3^{(2)}(t)
=
e^{-Q(t,T)}
\left(a_3^{(1)}(T)-a_3^{(2)}(T)\right),
\]
so
\begin{equation}
\left|a_3^{(1)}(t)-a_3^{(2)}(t)\right|
\le
e^{-\delta(t-T)}
\left|a_3^{(1)}(T)-a_3^{(2)}(T)\right|,
\qquad t\ge T.
\end{equation}
Thus the post-threshold forward dynamics rapidly forget the particular finite choice of local graph.

Second, every finite solution converges to the same asymptotic value. Since
\[
q(t)\to q_\infty:=4p_{\max}-\kappa>0
\qquad \text{as } t\to+\infty,
\]
the coefficient equation approaches the constant-coefficient limit
\[
\dot{a}_3+q_\infty a_3=-K,
\]
whose equilibrium is
\[
a_3^{\mathrm{asy}}
=
-\frac{K}{4p_{\max}-\kappa}
<0.
\]
Consequently,
\begin{equation}
a_3(t)\to a_3^{\mathrm{asy}}
=
-\frac{K}{4p_{\max}-\kappa}
\qquad \text{as } t\to+\infty.
\end{equation}
In particular, regardless of the finite value assigned to $a_3(T)$, there exists a later time
$T_1\ge T$ such that
\[
a_3(t)<0
\qquad \text{for all } t\ge T_1.
\]
Thus the negativity of the cubic bending coefficient is not imposed as an assumption; it is an eventual
dynamical consequence of the post-threshold forward equation itself.

Substituting
\[
y=h(x,t)=a_3(t)x^3+O(x^5)
\]
back into the $x$-equation yields the reduced dynamics along the critical graph. Since
\[
Ky(x^2+y^2)=Ka_3(t)x^5+O(x^7),
\]
the leading-order reduced equation is
\begin{equation}
\dot{x}=\mu(t)x-b(t)x^5,
\end{equation}
where
\[
\mu(t)=p(t)-\frac{\kappa}{2},
\qquad
b(t)=-Ka_3(t).
\]
Because $a_3(t)$ becomes negative after a transient, $b(t)$ becomes positive on every sufficiently late
post-threshold interval. The reduced dynamics therefore acquire the structure of a quintic normal
form: linear growth in the critical direction is saturated not by a direct cubic restoring term in $x$,
but by a quintic term generated through the slaving of the strongly damped transverse variable.

This is the key geometric point. The Kerr nonlinearity does not directly stabilize the critical
coordinate. Instead, it twists the state into the transverse $y$-direction, where dissipation acts, and
that damped transverse response feeds back into the $x$-dynamics as the first nontrivial restoring
term. The resulting quintic equation should therefore be regarded as the leading-order reduced model
for the local post-threshold dynamics. The actual nonautonomous pitchfork interpretation is then
established in the next subsection by analyzing the exact forward dynamics of this reduced equation
on a sufficiently late post-threshold interval, where $\mu(t)>0$ and $b(t)>0$.

\subsection{Exact Solution of the Reduced Equation and Post-Threshold Moving Branches}

We now analyze the reduced scalar equation derived in Section~4.2 on a sufficiently late
post-threshold interval:
\[
\dot{x}=\mu(t)x-b(t)x^5,
\]
where
\[
\mu(t)=p(t)-\frac{\kappa}{2},
\qquad
b(t)=-Ka_3(t).
\]
By the analysis of Section~4.2, there exists a sufficiently late time $T$ such that
\[
\mu(t)\ge \delta>0,
\qquad
b(t)>0,
\qquad
t\ge T,
\]
and such that the local invariant-graph reduction remains valid on the neighborhood under
consideration. The purpose of this subsection is to extract the exact solution formula for the reduced
dynamics and to identify the two symmetric moving branches that organize the local post-threshold
behavior.

It is important to distinguish this analysis from a pullback construction based on the limit
$t_0\to -\infty$. For the logistic preparation ramp used in Section~4.1, the coefficient $\mu(t)$ is
negative in the remote past and positive in the remote future. Accordingly, the branch selection
relevant for state formation is most naturally described on a forward time interval after the transition,
rather than by a global pullback limit from the distant past.

Because $x=0$ is an invariant solution of the reduced equation and the vector field is locally Lipschitz,
the sign of $x(t)$ is preserved along trajectories: positive initial data remain positive, and negative
initial data remain negative. On either sign half-line, introduce the transformation
\begin{equation}
w=x^{-4}.
\end{equation}
This converts the reduced equation into the linear nonautonomous equation
\begin{equation}
\dot{w}=-4\mu(t)w+4b(t).
\end{equation}
Given initial data $w(T)=w_T>0$, the exact solution is
\begin{equation}
w(t;T,w_T)
=
w_T e^{-4\int_T^t \mu(s)\,ds}
+
\int_T^t 4b(\tau)e^{-4\int_\tau^t \mu(s)\,ds}\,d\tau,
\qquad t\ge T.
\end{equation}

This formula yields several immediate consequences. First, because $\mu(t)\ge\delta>0$ for
$t\ge T$, the dependence on the initial value decays exponentially:
\[
|w(t;T,w_{T,1})-w(t;T,w_{T,2})|
=
|w_{T,1}-w_{T,2}|\,e^{-4\int_T^t \mu(s)\,ds}
\le
|w_{T,1}-w_{T,2}|\,e^{-4\delta(t-T)}.
\]
Thus, on the positive half-line, all $w$-solutions converge exponentially toward one another; the same
is true on the negative half-line by symmetry. Since the post-threshold solutions considered here
remain positive and bounded away from zero on every sufficiently late finite-time interval, this
exponential convergence of $w$ induces convergence of the corresponding amplitudes
$x=\pm w^{-1/4}$. The exact initial value affects only the transient phase lag, not the existence of
the two sign-separated attracting families.

It is therefore convenient to fix one representative positive solution by choosing any reference value
$w_*>0$ at time $T$, and to denote the corresponding symmetric pair of exact solutions by
\begin{equation}
\pm \rho(t)
=
\pm
\left[
w_* e^{-4\int_T^t \mu(s)\,ds}
+
\int_T^t 4b(\tau)e^{-4\int_\tau^t \mu(s)\,ds}\,d\tau
\right]^{-1/4},
\qquad t\ge T.
\end{equation}
These are exact forward-tracking solutions of the reduced equation. Different choices of $w_*>0$
produce different representatives of the same positive and negative attracting families, but all such
choices converge exponentially toward one another as $t\to+\infty$. Likewise, different admissible
finite choices of the local invariant graph in Section~4.2 affect the coefficient $b(t)$ only through
exponentially decaying transients. Hence the moving branches defined here are asymptotically
insensitive to the particular post-threshold initialization of the reduced graph.

The long-time behavior is determined by the asymptotic coefficients. Since
\[
\mu(t)\to \mu_\infty:=p_{\max}-\frac{\kappa}{2}>0,
\qquad
b(t)\to b_\infty>0,
\qquad
t\to+\infty,
\]
the linear equation for $w$ has the asymptotic equilibrium value
\[
w_\infty=\frac{b_\infty}{\mu_\infty},
\]
and hence
\[
\rho(t)\to
\left(\frac{\mu_\infty}{b_\infty}\right)^{1/4}
\qquad \text{as } t\to+\infty.
\]
Thus the reduced dynamics admit two symmetric nonzero asymptotic states, one on each sign half-line.

This yields the appropriate nonautonomous pitchfork interpretation of the reduced model. Before the
transition, the origin is the only locally attracting state. After the threshold has been crossed and the
reduction is valid, the origin becomes repelling within the reduced forward dynamics, while the local
phase space separates into two symmetric forward-attracting sign classes represented by the moving
branches $\pm \rho(t)$. In this sense, the reduced quintic equation provides the local dynamical model
for the birth and stabilization of the two logical states during the preparation protocol.

\subsection{Phase-Twist Stabilization and Hardware Implications}

The emergence of a quintic, rather than cubic, reduced normal form is a notable feature of the
resonant Kerr-cat geometry. In many symmetry-constrained oscillator models, cubic terms vanish for
structural reasons. Here the quintic term has a more specific interpretation: within the reduced
resonant semiclassical model, the critical coordinate is not stabilized directly by a cubic restoring
force. Instead, stabilization is generated indirectly through the transverse dissipative response.

The mechanism is encoded in the local invariant graph
\[
y\approx a_3(t)x^3.
\]
As the pump drives the state outward along the critical $x$-direction, the Kerr nonlinearity twists the
motion into the transverse $y$-direction. That induced $y$-component is strongly damped by the
single-photon loss, and the resulting slaved response feeds back into the $x$-equation through the
cross-term $Kyx^2$. Because the transverse response begins at cubic order in $x$, the leading restoring
contribution to the critical dynamics appears only at quintic order. This is the geometric origin of
the reduced equation
\[
\dot{x}=\mu(t)x-b(t)x^5.
\]

This interpretation also clarifies the physical role of dissipation. In the present local reduced
description, transverse dissipation is not merely a small perturbation of an otherwise autonomous
stabilizing mechanism. Rather, on sufficiently late post-threshold intervals it participates essentially
in the leading-order formation and stabilization of the local logical branches. Consequently, attempts
to heavily modify or suppress that transverse dissipative channel may alter the local state-formation
dynamics and the robustness of the resulting logical branches. This should be read as a model-based
design implication of the reduced resonant semiclassical picture, not as a universal statement about
all Kerr-cat implementations.

\begin{figure}[!htbp]
\centering
\includegraphics[width=0.94\textwidth]{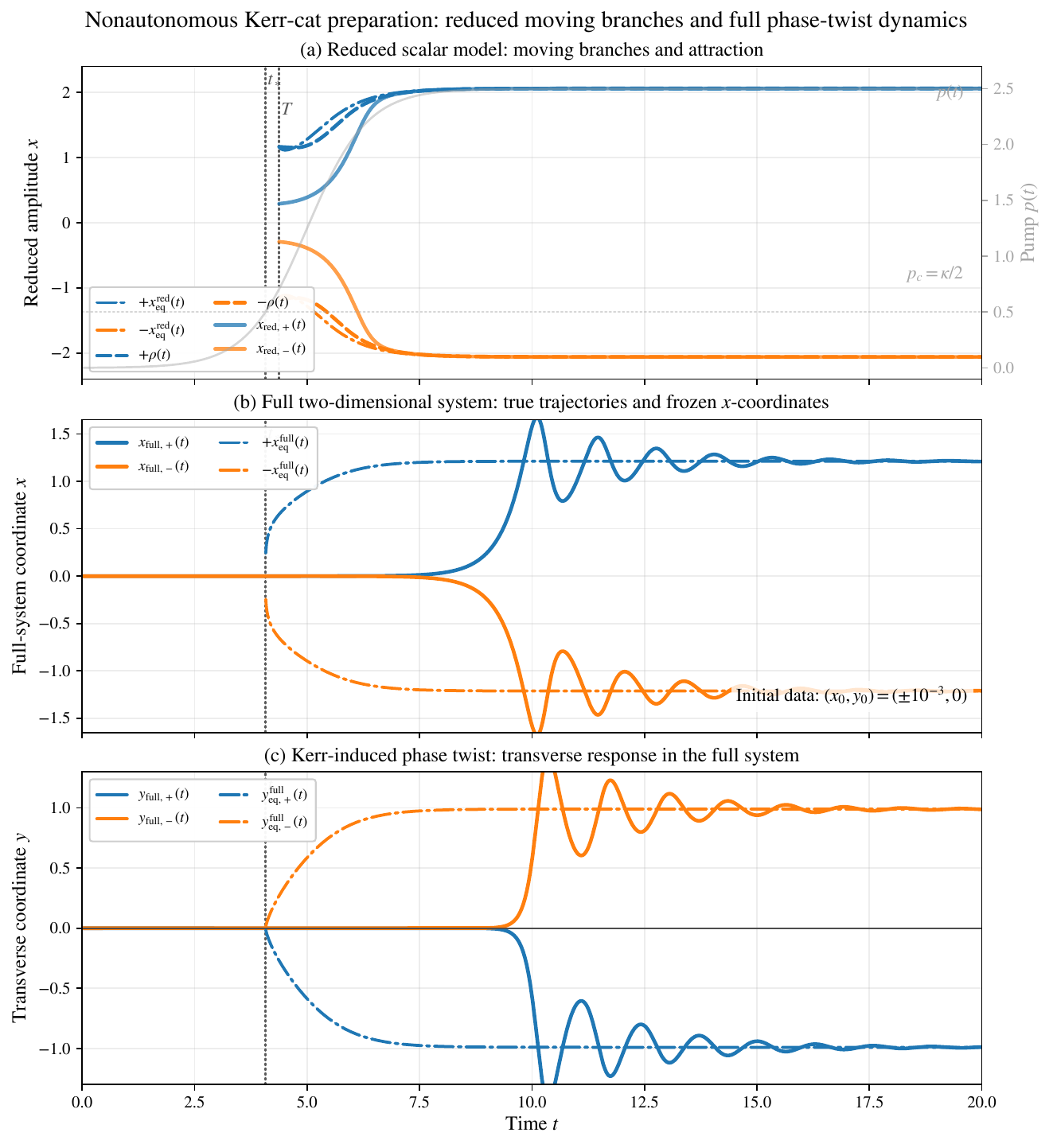}
\caption{Nonautonomous preparation dynamics separated into reduced and full-system diagnostics. 
(a) Reduced scalar model derived from the local invariant-graph reduction. The curves 
$\pm x^{\rm red}_{\rm eq}(t)$ are frozen-time algebraic roots of the reduced equation, 
$\pm \rho(t)$ are representative moving branches, and $x_{\rm red,\pm}(t)$ are reduced 
trajectories showing attraction toward those branches after the post-threshold reference time 
$T$. The gray curve shows the pump envelope $p(t)$, and the horizontal gray line marks 
$p_c=\kappa/2$. 
(b) Full two-dimensional system: $x$-coordinates of the true trajectories initialized at 
$(x_0,y_0)=(\pm 10^{-3},0)$, compared with the frozen-time full-system equilibrium 
coordinates. 
(c) Full two-dimensional system: transverse $y$-coordinates of the same trajectories, 
showing the Kerr-induced phase twist and dissipative relaxation toward the frozen full-system 
branches. Panel (a) displays the local reduced post-threshold mechanism, whereas panels 
(b) and (c) display two coordinate projections of the same full hardware-coordinate 
trajectories; the similar damped envelopes in panels (b) and (c) reflect the same 
Kerr-induced phase-twist relaxation, in which the transverse response is dissipatively 
slaved back into the $x$-dynamics. The reduced and full-system curves are not asserted to 
coincide quantitatively at large amplitude. Numerical parameters are listed in Appendix A.}
\label{fig:adiabatic_lag}
\end{figure}

Figure~\ref{fig:adiabatic_lag} separates the preparation dynamics into reduced and full-system
diagnostics. Panel~(a) concerns only the reduced scalar equation derived from the local invariant-graph
calculation. It compares the frozen-time algebraic roots of the reduced equation with representative
moving branches and with reduced scalar trajectories. The reduced moving branches are true solutions
of the nonautonomous reduced equation, while the algebraic roots are instantaneous zeros of the
reduced vector field; after the pump saturates, both approach the same limiting autonomous
amplitudes.

Panels~(b) and~(c) then show the corresponding full two-dimensional hardware dynamics. The two
full-system trajectories are initialized symmetrically near the vacuum, at
\[
(x_0,y_0)=(\pm 10^{-3},0),
\]
and are compared with the frozen-time nonzero equilibria of the full planar system. These full-system
trajectories need not coincide quantitatively with the reduced moving branches at large amplitude:
the reduced equation is a local post-threshold model near the critical direction, whereas the full
trajectories evolve in the complete two-dimensional phase space. The visible overshoot and damped
oscillation in panel~(b), together with the transverse response in panel~(c), illustrate the Kerr-induced
phase twist described above: growth in the critical coordinate generates a transverse $y$-component,
which is then dissipatively slaved back into the $x$-dynamics.

The supplementary numerical results in Appendix~B.1 reinforce this interpretation in two ways.
First, increasing the ramp rate amplifies the lag between the true trajectory and the instantaneous
algebraic roots, illustrating, for the parameter values shown, how frozen-time equilibrium pictures can
underestimate the non-adiabaticity of fast preparation protocols. Second, different finite post-threshold
initializations of the reduced moving branch converge rapidly toward one another, supporting the
robustness of the branch construction even though the local invariant graph is not selected by a unique
canonical initial condition on the whole time axis.

Thus the preparation dynamics are governed not by a static one-dimensional potential picture, but
by a genuinely two-dimensional nonautonomous phase-space mechanism: Kerr-induced twisting into
the transverse direction, dissipation acting on that transverse response, and delayed feedback into the
critical coordinate. The resulting quintic reduced equation is therefore best interpreted as the leading
local dynamical model for the birth and stabilization of the logical branches during nonautonomous
state preparation.

\section{Gate Operations and Phase Space Transport}

Once the logical states have been created by the preparation protocol, gate execution is modeled as
a localized perturbation of the nearby operating configuration. The preparation analysis in
Section~4 describes the birth of the two logical branches under a nonautonomous ramp. By contrast,
the purpose of the present section is to analyze the onset of transport during a \emph{fast gate pulse}
after the system has entered the operating regime.

For this purpose, we approximate the gate dynamics as a weak, aperiodically time-dependent
perturbation of the conservative resonant skeleton constructed in Section~3. This modeling step is
natural from the point of view of phase-space transport. The unperturbed Hamiltonian system
provides a figure-eight separatrix that sharply divides the two logical regions. A fast gate pulse does
not merely deform this boundary quasistatically. Rather, once time dependence is introduced, the
stable and unstable curves associated with the saddle need no longer coincide. Their splitting creates
transient lobe structures, and these lobes provide a geometric mechanism for non-adiabatic transport
between the two logical regions.

\subsection{The Perturbed Vector Field and the Melnikov Function}

We work in the parametrically resonant case $\Delta=0$ for the unperturbed skeleton and fix a
constant baseline pump $p_0>0$. This yields the bounded figure-eight homoclinic separatrix
parameterized by $(x_h(t),y_h(t))$ from Section~3. We then model dissipation, small detuning, and
the transient gate envelope as weak perturbations of this integrable Hamiltonian configuration.

Let $0<\epsilon\ll 1$ be a formal bookkeeping parameter to scale the perturbation. This gate
calculation uses a different time dependence from the preparation ramp. After preparation, we
expand about a constant operating pump $p_0$ and superimpose a localized pulse. To avoid confusion
with the preparation ramp, denote the total gate-stage pump by
\[
P(t)=p_0+\epsilon p_1(t),
\qquad
p_1(t)=A e^{-t^2/(2\sigma^2)},
\]
where $p_1(t)$ is the localized gate envelope, with amplitude $A>0$ and width $\sigma>0$. The
Gaussian pulse decays rapidly as $t\to\pm\infty$, which ensures convergence of the Melnikov
integrals. We also scale the dissipation and detuning as $\kappa\mapsto \epsilon\kappa$ and
$\Delta\mapsto \epsilon\Delta$. Here $\epsilon$ serves as a formal ordering parameter for the
analytical derivation. The numerical illustrations are computed with the physical parameter values
directly, in a regime where the first-order Melnikov approximation remains informative.

Then the planar vector field, with the preparation-stage pump replaced by the gate-stage pump
$P(t)$, can be written in the perturbed Hamiltonian form
\begin{equation}\label{eq:perturbed_vector_field}
\begin{pmatrix}
\dot{x}\\
\dot{y}
\end{pmatrix}
=
\begin{pmatrix}
\partial H/\partial y\\
-\partial H/\partial x
\end{pmatrix}
+
\epsilon
\begin{pmatrix}
p_1(t)x-\dfrac{\kappa}{2}x+\Delta y\\[4pt]
-p_1(t)y-\dfrac{\kappa}{2}y-\Delta x
\end{pmatrix}.
\end{equation}
The perturbation term isolates the three physically relevant effects: nonautonomous gate modulation,
single-photon loss, and small detuning.

Under this perturbation, the coincident stable and unstable manifolds of the unperturbed saddle
split into distinct time-dependent invariant objects. In this planar problem these manifolds are
one-dimensional curves. To measure their leading-order separation, we evaluate
the aperiodic Melnikov function associated with the homoclinic orbit. For a cross-section time
$t_0$, the Melnikov function is
\begin{equation}\label{eq:melnikov_def}
M(t_0)
=
\int_{-\infty}^{\infty}
\left[
\dot{x}_h(t)\,g_y(x_h(t),y_h(t),t+t_0)
-
\dot{y}_h(t)\,g_x(x_h(t),y_h(t),t+t_0)
\right]\,dt,
\end{equation}
where $g=(g_x,g_y)$ denotes the perturbation vector field. Substituting the perturbation components
from Equation~\eqref{eq:perturbed_vector_field} yields
\begin{equation}\label{eq:melnikov_split}
\begin{aligned}
M(t_0)
&=
\int_{-\infty}^{\infty}
-p_1(t+t_0)\bigl(x_h\dot{y}_h+y_h\dot{x}_h\bigr)\,dt  \\[4pt]
&\qquad
+
\frac{\kappa}{2}
\int_{-\infty}^{\infty}
\bigl(x_h\dot{y}_h-y_h\dot{x}_h\bigr)\,dt
-
\Delta
\int_{-\infty}^{\infty}
\bigl(x_h\dot{x}_h+y_h\dot{y}_h\bigr)\,dt .
\end{aligned}
\end{equation}

Each term has a clear geometric meaning. For the gate modulation term, an integration by parts
moves the derivative onto the pulse profile:
\[
-\int_{-\infty}^{\infty}
p_1(t+t_0)\bigl(x_h\dot{y}_h+y_h\dot{x}_h\bigr)\,dt
=
\int_{-\infty}^{\infty}
\dot{p}_1(t+t_0)x_h(t)y_h(t)\,dt,
\]
where the boundary term vanishes because the homoclinic orbit approaches the saddle as
$t\to\pm\infty$. Thus manifold splitting is driven by the temporal ramping of the gate rather than
by the pulse amplitude alone.

For the dissipation term, the signed area integral is orientation-dependent. We write
\[
\frac{1}{2}
\int_{-\infty}^{\infty}
\bigl(x_h\dot{y}_h-y_h\dot{x}_h\bigr)\,dt
=
-A_{\mathrm{lobe}},
\qquad
A_{\mathrm{lobe}}>0,
\]
where $A_{\mathrm{lobe}}$ is the positive geometric area enclosed by the unperturbed lobe. Equivalently,
the homoclinic branch has been oriented so that the signed area integral is negative. With this
orientation convention, the dissipative contribution to the Melnikov function is
\[
\frac{\kappa}{2}
\int_{-\infty}^{\infty}
\bigl(x_h\dot{y}_h-y_h\dot{x}_h\bigr)\,dt
=
-\kappa A_{\mathrm{lobe}}.
\]
This gives the constant negative offset that opposes pulse-induced splitting. If the opposite
orientation of the homoclinic orbit is used, the Melnikov function is multiplied by an overall sign;
the zero condition for transversal intersections is unchanged, provided the convention is used
consistently.

For the symmetric resonant separatrix used as the unperturbed orbit, the detuning contribution
reduces to a total derivative and therefore vanishes at first order. Indeed,
\[
M_\Delta
=
-\Delta
\int_{-\infty}^{\infty}
\bigl(x_h\dot{x}_h+y_h\dot{y}_h\bigr)\,dt
=
-\frac{\Delta}{2}
\int_{-\infty}^{\infty}
\frac{d}{dt}\bigl(x_h(t)^2+y_h(t)^2\bigr)\,dt,
\]
so that
\begin{equation}\label{eq:melnikov_detuning_zero}
M_\Delta
=
-\frac{\Delta}{2}
\left[
x_h(t)^2+y_h(t)^2
\right]_{-\infty}^{\infty}
=
0,
\end{equation}
since the homoclinic orbit begins and ends at the saddle point at the origin. This cancellation is
specific to the first-order Melnikov calculation about the resonant skeleton; it should not be
interpreted as a universal detuning-protection theorem. Higher-order terms may of course reintroduce
detuning effects, but they are beyond the scope of the present first-order analysis.

The simplified Melnikov function is therefore
\begin{equation}\label{eq:melnikov_final}
M(t_0)
=
\int_{-\infty}^{\infty}
\dot{p}_1(t+t_0)x_h(t)y_h(t)\,dt
-
\kappa A_{\mathrm{lobe}}.
\end{equation}
Equation~\eqref{eq:melnikov_final} expresses the central competition governing gate-induced
transport: the pulse ramp tends to split the invariant curves and create transport lobes, while
dissipation produces a constant negative offset that opposes their intersection.

\begin{figure}[!htbp]
    \centering
    \includegraphics[width=0.85\textwidth]{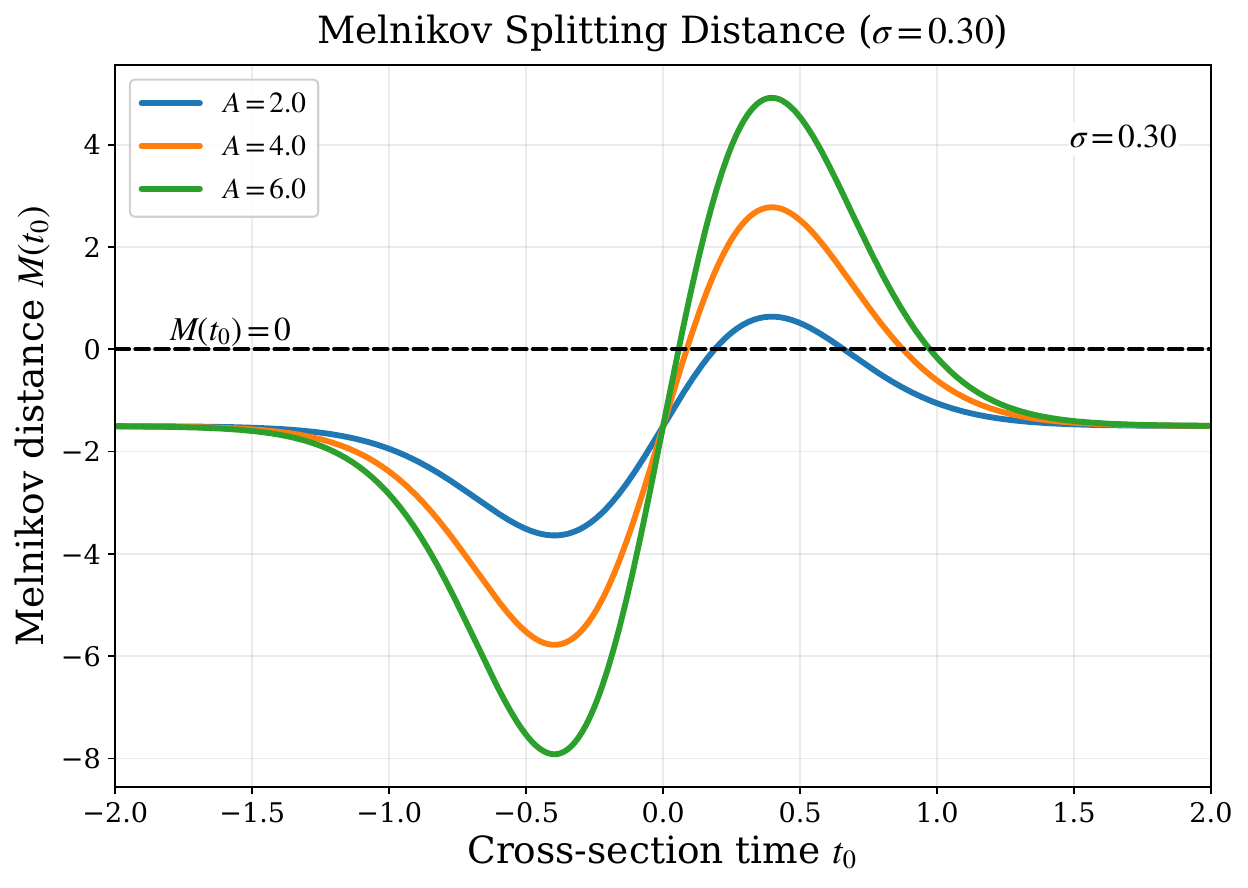}
\caption{\small Signed leading-order Melnikov splitting distance $M(t_0)$ for a Gaussian gate
    pulse with fixed width $\sigma=0.30$ and amplitudes $A=2.0,4.0,6.0$, computed from the
    analytic resonant homoclinic branch used in the canonical Melnikov calculation. The dashed
    horizontal line marks $M(t_0)=0$. For amplitudes large enough that $M(t_0)$ has simple zeros,
    the time-dependent stable and unstable curves intersect transversally, indicating the onset of
    transient lobe-mediated transport. The negative offset away from the pulse region is produced by
    dissipation.}
    \label{fig:melnikov_zeros}
\end{figure}

\subsection{Transient Lobe Dynamics and Leading-Order Leakage Thresholds}

Equation~\eqref{eq:melnikov_final} provides a leading-order criterion for the onset of transport in
the semiclassical phase-space model. In the context of phase-space reaction dynamics, the
unperturbed stable manifold of the saddle acts as a separatrix, forming a nominal
\textit{dividing surface} (here, the $x=0$ axis near the origin) that separates the two logical regions.
In the absence of intersections, this dividing surface operates as an intact transport barrier preventing
non-adiabatic crossings.

When $M(t_0)$ has a simple zero,
\[
M(t_0)=0,
\qquad
M'(t_0)\neq 0,
\]
the time-dependent stable and unstable curves intersect transversally. In the present aperiodic
setting, such intersections do not generate a permanent Smale horseshoe. Instead, they produce a
\emph{transient homoclinic tangle} with a finite number of lobes, in the sense of Malhotra and
Wiggins~\cite{Malhotra1998}. These lobes act as a geometric turnstile across the dividing surface and
mediate a finite burst of phase-space transport. This is the mechanism we associate with
non-adiabatic leakage during a fast gate pulse. Once transversal intersections occur, the barrier is
replaced temporarily by a corrugated lobe structure, and classical phase-space area can be transported
across the dividing boundary during the pulse.

To organize the threshold in parameter space, we regard the Melnikov function as depending on the
gate amplitude $A$ and pulse width $\sigma$, and define
\begin{equation}\label{eq:mmax_def}
M_{\max}(A,\sigma)
=
\max_{t_0\in\mathbb{R}} M(t_0;A,\sigma).
\end{equation}
The leading-order threshold between a transport-inactive regime and a transport-active regime is
then given by the zero level set
\[
M_{\max}(A,\sigma)=0.
\]
When $M_{\max}(A,\sigma)<0$, dissipation dominates the pulse-induced splitting across all cross
sections, and the invariant curves remain separated. When $M_{\max}(A,\sigma)>0$, the Melnikov
function has at least one simple zero, indicating the onset of transversal intersections and transient
lobe-mediated transport.

It is important to emphasize the scope of this criterion. The curve $M_{\max}(A,\sigma)=0$ is not a
sharp quantum gate-fidelity boundary; rather, it is a leading-order geometric onset threshold for
classical transport in the semiclassical model. For a Gaussian pulse, the characteristic control rate
scales like $A/\sigma$, so this threshold constrains how rapidly and strongly the pump can be
modulated before transient lobe transport becomes available. Its value lies in identifying where fast
pulses first become capable of breaking the nominal transport barrier created by the resonant
separatrix.

\begin{figure}[!htbp]
    \centering
    \includegraphics[width=0.85\textwidth]{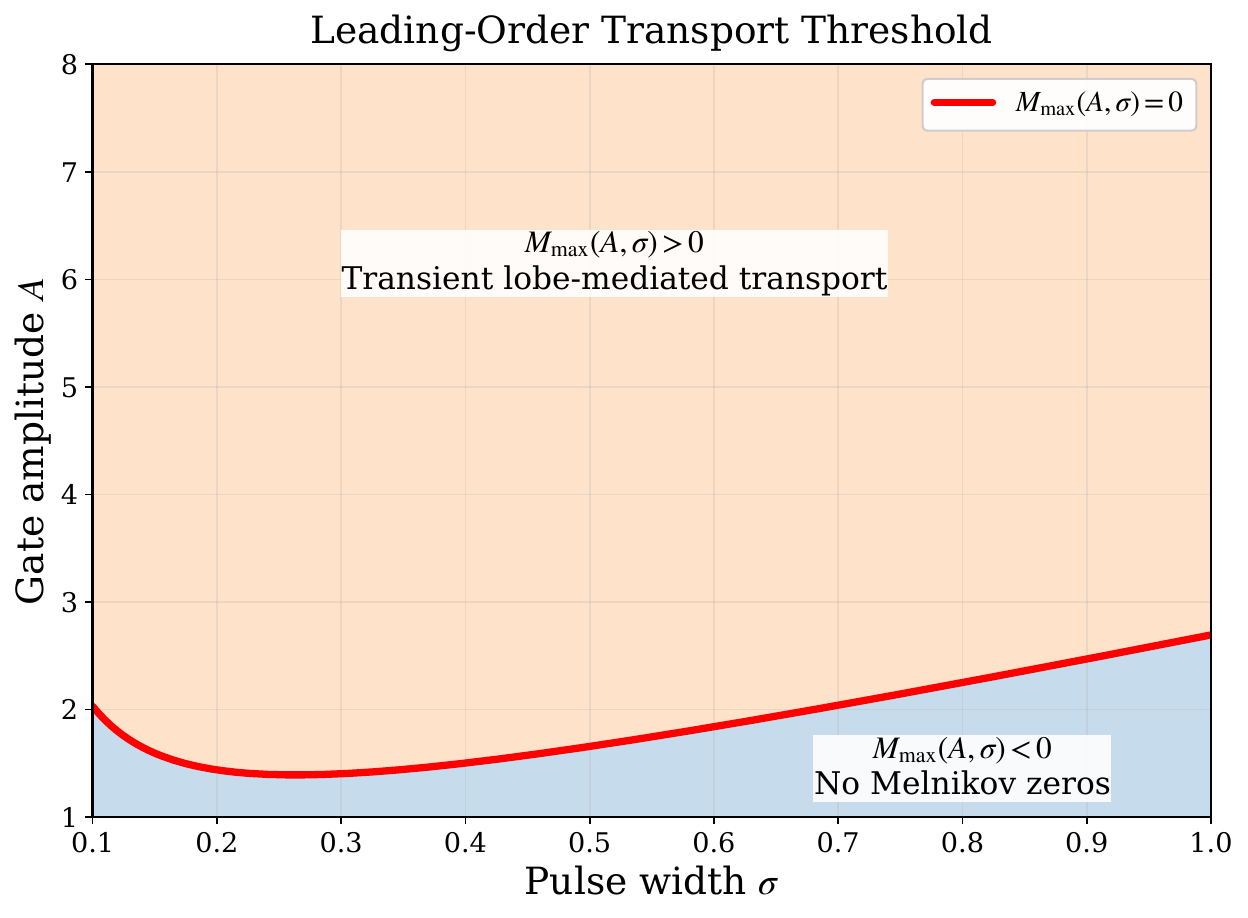}
    \caption{\small Leading-order transport threshold for gate-induced leakage in the semiclassical
    model. The red curve is the canonical numerically evaluated level set $M_{\max}(A,\sigma)=0$
    in the gate amplitude-width plane, plotted with pulse width $\sigma$ on the horizontal axis and
    gate amplitude $A$ on the vertical axis. Below the curve, $M_{\max}(A,\sigma)<0$, no Melnikov
    zeros are detected, and the first-order calculation predicts no transversal intersections of the
    time-dependent stable and unstable curves. Above the curve, $M_{\max}(A,\sigma)>0$, the
    aperiodic pulse produces transient transversal intersections, and transient lobe-mediated transport
    across the nominal dividing boundary becomes available. This is a leading-order geometric onset
    criterion, not a sharp quantum gate-fidelity boundary.}
    \label{fig:transport_threshold}
\end{figure}

This interpretation also clarifies the meaning of Figure~\ref{fig:transport_threshold}. The
region below the red curve corresponds to the condition that no Melnikov zeros are detected in the
first-order calculation. The region above the curve should not be interpreted as a permanent chaotic
phase in the strict Smale-horseshoe sense. Rather, it is a regime in which the aperiodic pulse
produces transient intersections and hence transient lobe-mediated transport. The corresponding
phase-space transport provides a natural semiclassical mechanism for non-adiabatic leakage and
resulting logical bit-flips during gate execution. An illustrative full-system transport simulation
across the dividing surface is provided in Appendix~B.2.

\FloatBarrier

\section{Discussion and Outlook}

The analysis in this paper has been deliberately organized around two related but distinct
nonautonomous problems. Section~4 addresses state preparation under a ramped control pulse, while
Section~5 analyzes gate-induced transport by perturbing the conservative resonant operating
skeleton. These two analyses serve different purposes: the first isolates the local birth and
stabilization of the logical branches, and the second identifies a leading-order transport threshold
during fast gate execution. Both are semiclassical, and both use the parametrically resonant geometry
as the organizing limit. In Section~5, small detuning is included as a perturbation and shown to
cancel from the first-order Melnikov splitting for the symmetric resonant skeleton.

Within this framework, the main lesson is that a purely frozen-time picture is insufficient. During
preparation, the relevant reduced dynamics are not governed by a static one-dimensional cubic
potential. Instead, the local nonautonomous invariant-graph reduction yields a quintic normal form
in the critical coordinate. The first restoring term appears only after the Kerr-induced transverse
response is slaved to the dissipative direction and fed back into the critical dynamics. In the
post-threshold regime, this reduced equation supports two symmetric moving branches that organize
local state formation and make the geometric origin of non-adiabatic lag explicit.

For gate execution, the Melnikov calculation gives a complementary message. Fast aperiodic pulses
should not be viewed merely as quasistatic deformations of an otherwise intact barrier. At leading
order, they can split the time-dependent stable and unstable curves associated with the resonant
separatrix and generate transient lobe dynamics. In the semiclassical model, the threshold curve
\[
M_{\max}(A,\sigma)=0
\]
is therefore best interpreted as a leading-order geometric onset criterion for transport, not as a sharp
quantum gate-fidelity boundary. Likewise, the first-order cancellation of the detuning contribution
should be read as a property of the symmetric resonant Melnikov skeleton rather than as a universal
detuning-protection theorem.

Several limitations should be kept in view. The invariant-graph reduction in Section~4 is a formal
local reduction near the vacuum trajectory, not a global center-manifold theorem on all of
\(\mathbb{R}\). The post-threshold moving branches are extracted from the reduced scalar equation on
a forward time interval after the transition, rather than from a global pullback construction. The
transport threshold in Section~5 is first-order in the perturbation size and is derived for a specific
class of localized pulses. Higher-order corrections, stronger detuning effects, nonresonant
preparation dynamics, and deeper quantum-regime phenomena may modify the precise location of the
onset curves.

Even with these caveats, the geometric picture that emerges is useful. The manuscript identifies a
coherent phase-space description of how logical states form under nonautonomous preparation and
how those states become vulnerable to transport under fast pulses. This suggests several natural
continuations: a more systematic nonautonomous bifurcation treatment of the reduced preparation
problem; a quantum-semiclassical comparison of lobe-mediated transport with leakage measured from
Wigner-function evolution; and an extension from one-dimensional dividing curves to
higher-dimensional time-dependent NHIMs in coupled Kerr-cat architectures. Each of these directions
would sharpen the connection between nonautonomous phase-space geometry and the practical limits
of continuous-variable quantum hardware.

\appendix

\renewcommand{\thefigure}{\thesection\arabic{figure}}
\renewcommand{\theHfigure}{\thesection\arabic{figure}}
\makeatletter
\@addtoreset{figure}{section}
\makeatother

\section{Numerical Methods and Reproducibility}

To support reproducibility, we record the parameter values and numerical procedures used for the
state-preparation and gate-transport calculations. All time integrations were performed with a
Backward Differentiation Formula (BDF) solver in order to handle the stiffness induced by the
dissipative terms. Relative tolerances were set to $10^{-8}$ and absolute tolerances to $10^{-10}$.

\begin{table}[!htbp]
\centering
\begin{tabular}{lll}
\toprule
Symbol & Physical Parameter / Numerical Setting & Value \\
\midrule
$\kappa$ & Single-photon dissipation rate & $1.0$ \\
$K$ & Kerr nonlinearity constant & $1.0$ \\
$p_0$ & Baseline parametric pump amplitude & $1.5$ \\
$p_{\max}$ & Maximum logistic ramp amplitude & $2.5$ \\
$\gamma$ & Logistic ramp rate & $1.5$ \\
$t_c$ & Logistic temporal center & $5.0$ \\
$\sigma$ & Gaussian gate pulse width sweep domain & $[0.1,1.0]$ \\
$A$ & Gaussian gate pulse amplitude sweep domain & $[1.0,8.0]$ \\
\bottomrule
\end{tabular}
\caption{Parameter values used for the numerical integrations and Melnikov evaluations.}
\label{tab:numerical_parameters}
\end{table}

\FloatBarrier

\subsection{State Initialization and Forward-Tracking Branches}

Figure~\ref{fig:adiabatic_lag} separates the preparation calculation into two related diagnostics. The
first is the reduced scalar model obtained from the local invariant-graph reduction in Section~4. The
second is the full two-dimensional nonautonomous hardware model in the original $(x,y)$ coordinates.
This distinction is important: the reduced branches organize the local post-threshold mechanism near
the critical direction, whereas the full trajectories display the subsequent two-dimensional
phase-twist relaxation.

For panel~(a), the coefficient equation
\[
\dot a_3=-K-(4p(t)-\kappa)a_3
\]
was integrated forward from a sufficiently late post-threshold reference time $T$. The transformed
linear equation
\[
\dot w
=
-4\left(p(t)-\frac{\kappa}{2}\right)w+4b(t),
\qquad
b(t)=-Ka_3(t),
\]
was then integrated forward from a fixed positive initial value $w(T)=w_*>0$. The representative
positive and negative moving branches were defined by
\[
\pm\rho(t)=\pm w(t)^{-1/4}.
\]
The reduced scalar trajectories $x_{\mathrm{red},\pm}(t)$ shown in panel~(a) were initialized away from
these representative branches in order to illustrate attraction within the reduced equation. The
instantaneous algebraic roots of the reduced model were computed from
\[
x_{\mathrm{eq}}^{\mathrm{red}}(t)
=
\pm\left(\frac{\mu(t)}{b(t)}\right)^{1/4},
\qquad
\mu(t)=p(t)-\frac{\kappa}{2},
\]
whenever $\mu(t)>0$ and $b(t)>0$. These algebraic roots are not trajectories of the nonautonomous
reduced equation; they are frozen-time reference states. Since the logistic pump saturates and the
coefficients approach constants, the reduced moving branches and the algebraic roots approach the
same limiting autonomous amplitudes at late times.

For panels~(b) and~(c), the full hardware trajectories were obtained by integrating the planar
nonautonomous vector field given by Equations~\eqref{eq:x_dot} and~\eqref{eq:y_dot} forward in time from
the symmetric initial conditions
\[
(x_0,y_0)=(\pm 10^{-3},0).
\]
Panel~(b) plots the critical coordinate $x(t)$ for these two full-system trajectories. Panel~(c) plots
the corresponding transverse coordinate $y(t)$, which displays the Kerr-induced phase twist and its
subsequent dissipative relaxation.

For comparison, panels~(b) and~(c) also show the frozen-time nonzero equilibria of the full
resonant planar system. For $\Delta=0$ and $p(t)>p_c=\kappa/2$, these frozen equilibria satisfy
\[
K^2 r^4=p(t)^2-\left(\frac{\kappa}{2}\right)^2,
\qquad
r^2=x^2+y^2,
\]
and
\[
y=-\frac{p(t)-\kappa/2}{K r^2}\;x,
\]
with the sign of $x$ selecting the positive or negative branch. These frozen full-system equilibria are
included only as instantaneous reference curves. They are not trajectories of the nonautonomous
system. The damped oscillations of the true full-system trajectories in panels~(b) and~(c) are the
visible signature of the two-dimensional phase-twist mechanism discussed in Section~4.4.

\subsection{Melnikov Integrals and Leading-Order Transport Thresholds}

The Melnikov calculations were carried out for the resonant conservative skeleton
\[
\Delta = 0,\qquad \kappa = 0,\qquad p(t)=p_0,
\]
for which Section~3 yields a bounded figure-eight homoclinic separatrix. The positive geometric
area enclosed by one lobe is
\[
A_{\mathrm{lobe}}=\frac{4p_0}{3K}.
\]
This is a geometric area and is therefore positive. In the Melnikov calculation, however, the area
appears through the oriented line integral
\[
\frac{1}{2}
\int_{-\infty}^{\infty}
\bigl(x_h\dot y_h-y_h\dot x_h\bigr)\,dt.
\]
With the homoclinic orientation used in Section~5, this signed integral is
\[
\frac{1}{2}
\int_{-\infty}^{\infty}
\bigl(x_h\dot y_h-y_h\dot x_h\bigr)\,dt
=
-A_{\mathrm{lobe}},
\qquad
A_{\mathrm{lobe}}>0.
\]
Thus the dissipative contribution to the Melnikov function is the negative offset
$-\kappa A_{\mathrm{lobe}}$. Reversing the orientation of the homoclinic parameterization would
multiply the Melnikov function by an overall sign, but would not change the zero condition for
transversal intersections.

The unperturbed homoclinic orbit $(x_h(t),y_h(t))$ entering the canonical Melnikov calculation
was evaluated using the analytic resonant parameterization of one oriented branch of the $H=0$
separatrix,
\[
x_h(t)=\sqrt{\frac{p_0}{K}}\,e^{p_0 t}\operatorname{sech}(2p_0t),
\qquad
 y_h(t)=-\sqrt{\frac{p_0}{K}}\,e^{-p_0 t}\operatorname{sech}(2p_0t).
\]
This choice satisfies $H(x_h(t),y_h(t))=0$ and tends to the saddle at the origin as
$t\to\pm\infty$. It also fixes the orientation convention used above, so that the signed line
integral is negative and the dissipative Melnikov contribution is $-\kappa A_{\mathrm{lobe}}$.
The finite-window numerical evaluation used the interval $t\in[-20,20]$ with a uniform grid and
trapezoidal quadrature; the maximum numerical deviation of $H(x_h(t),y_h(t))$ from zero was used
as a check on the implementation.

For Figure~\ref{fig:melnikov_zeros}, the leading-order Melnikov function
\[
M(t_0)
=
\int_{-\infty}^{\infty}
\dot p_1(t+t_0)x_h(t)y_h(t)\,dt
-
\kappa A_{\mathrm{lobe}}
\]
was evaluated on the same homoclinic grid. The cross-section parameter was sampled over
$t_0\in[-2,2]$ using 801 grid points, and the integral was evaluated over $t\in[-20,20]$ using
4001 points and trapezoidal quadrature. The plotted curves use $\sigma=0.30$ and
$A=2.0,4.0,6.0$. The rapid decay of $\dot p_1(t+t_0)$ and of the homoclinic factor
$x_h(t)y_h(t)$ justifies the finite truncation.

For Figure~\ref{fig:transport_threshold}, the gate-induced transport threshold was extracted
from the scalar field
\[
M_{\max}(A,\sigma)
=
\max_{t_0\in\mathbb{R}}M(t_0;A,\sigma).
\]
Because the Gaussian pulse enters linearly in the amplitude $A$, the Melnikov function can be written
as
\[
M(t_0;A,\sigma)=A\,G_\sigma(t_0)-\kappa A_{\mathrm{lobe}},
\]
where
\[
G_\sigma(t_0)=\int_{-\infty}^{\infty}
\frac{d}{dt}\left(e^{-(t+t_0)^2/(2\sigma^2)}\right)x_h(t)y_h(t)\,dt.
\]
Thus the threshold satisfies
\[
A_{\mathrm{crit}}(\sigma)
=
\frac{\kappa A_{\mathrm{lobe}}}{\max_{t_0}G_\sigma(t_0)},
\]
whenever the denominator is positive. The canonical threshold curve was evaluated for
$\sigma\in[0.1,1.0]$ using 181 grid points; for each $\sigma$, $G_\sigma(t_0)$ was computed on
$t_0\in[-2,2]$ using the same homoclinic quadrature described above. The resulting curve is shown
with $\sigma$ on the horizontal axis and $A$ on the vertical axis.
This curve should be interpreted as a leading-order Melnikov threshold for the onset of transient
lobe-mediated transport in the semiclassical model. It is not a sharp quantum gate-fidelity boundary,
but rather a geometric criterion for when the aperiodic pulse first becomes capable of producing
transversal intersections of the time-dependent stable and unstable curves.

\FloatBarrier

\section{Supplementary Numerical Validations}

To further support the geometric thresholds and nonautonomous reductions presented in the main
text, we provide two supplementary numerical diagnostics targeting the preparation dynamics and the
gate execution limits.

\subsection{Ramp-Rate Dependence and Forward-Tracking Convergence}

\begin{figure}[!htbp]
    \centering
    \includegraphics[width=0.75\textwidth]{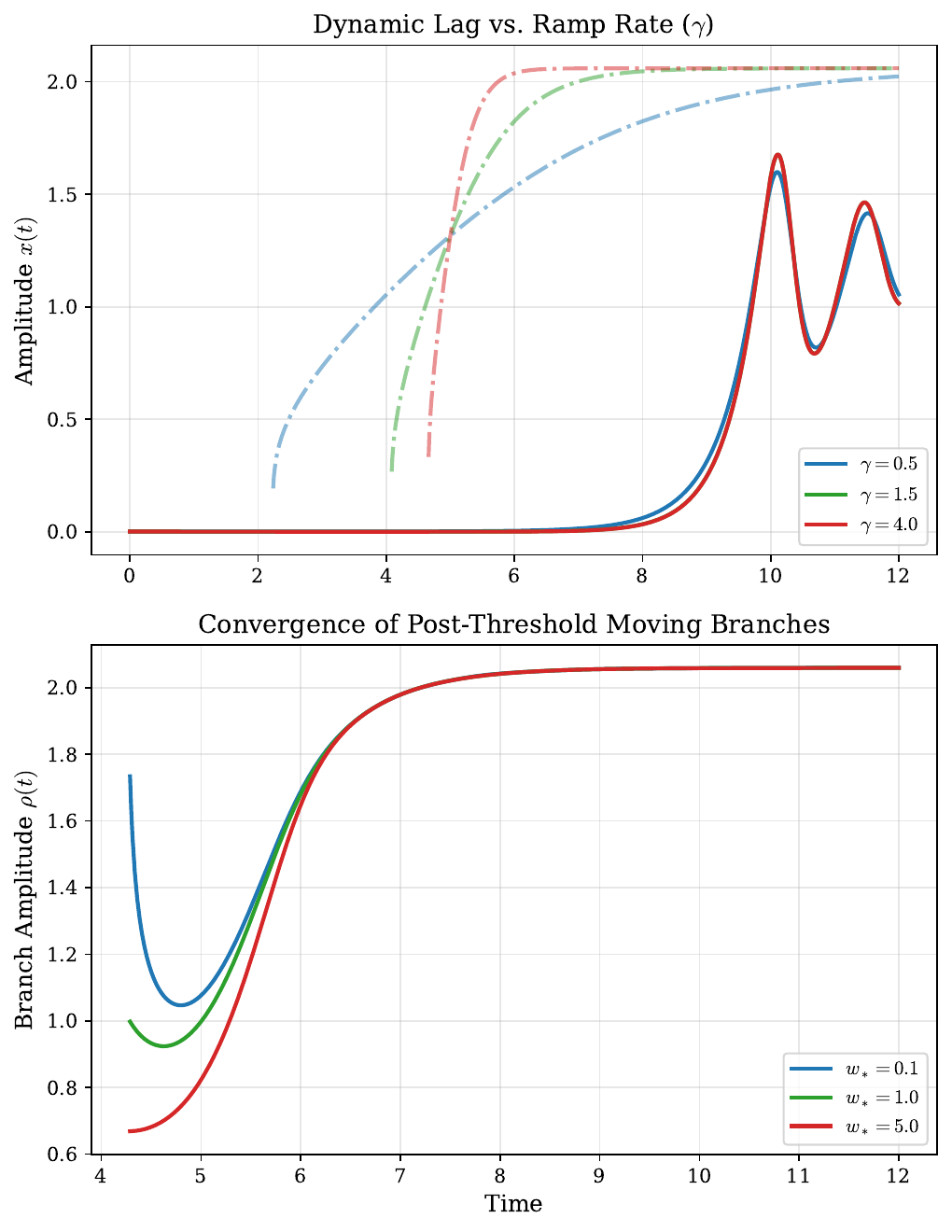}
    \caption{Supplementary preparation diagnostics. (Top) Dynamic lag between a true semiclassical
    trajectory and the instantaneous algebraic roots for several logistic ramp rates $\gamma$. Faster
    ramps produce larger lags, illustrating why frozen-time equilibria are unreliable for rapid
    preparation. (Bottom) Convergence of representative post-threshold moving branches $\rho(t)$
    initialized from different finite values $w_*$. The rapid convergence illustrates that the
    forward-tracking branch construction is insensitive to the particular post-threshold initialization,
    up to exponentially decaying transients.}
    \label{fig:appendix_ramprate}
\end{figure}

A central claim of Section~4 is that frozen-time algebraic equilibria fail to capture the true operational
location of the quantum hardware states during initialization. Figure~\ref{fig:appendix_ramprate}
(Top) illustrates the dynamic lag for three different logistic ramp rates
($\gamma\in\{0.5,1.5,4.0\}$). As the temporal ramping speed increases, the true physical trajectory
(solid lines) exhibits an increasingly severe geometric lag behind the instantaneous algebraic roots
(dash-dot lines). This illustrates, for the parameter values shown, how frozen-time equilibrium models
can underestimate the non-adiabaticity of fast preparation protocols.

Furthermore, Section~4.3 models the logical states using exact forward-tracking moving branches
$\pm\rho(t)$ initialized at a sufficiently late post-threshold reference time $T$.
Figure~\ref{fig:appendix_ramprate} (Bottom) plots the reduced branch amplitude evaluated from
three disparate initial boundary conditions $w_*$. As derived analytically, the differences in initial
data affect only exponentially decaying transients; all solutions rapidly converge onto the identical
moving branch. This illustrates the robustness of the forward-tracking branch interpretation.

\subsection{Full-System Simulation of Lobe-Mediated Transport}

Section~5 utilizes the aperiodic Melnikov method to extract a leading-order threshold
$M_{\max}(A,\sigma)=0$ for the onset of transient lobe-mediated transport
(Figure~\ref{fig:transport_threshold}). To test whether this geometric mechanism is reflected in
the full nonautonomous system, we simulate the transport lobe for a gate pulse operating in the active
regime $(A=7.5,\sigma=0.3)$.

Analytically, we predict the location of the escaping turnstile lobe by computing the simple zeros of
the Melnikov function $M(t_0)=0$. These zeros extract the leading-order cross-section times at which
the time-dependent stable and unstable manifolds intersect. By evaluating the unperturbed homoclinic
orbit at these times, specifically at the coordinates $(x_h(-t_0),y_h(-t_0))$, we project these
intersection points back onto the unperturbed separatrix to identify the geometric base of the
predicted transport lobe.

To reveal the true perturbed lobe dynamics, we generate a ring of 150 initial conditions centered in
the right logical region of the unperturbed phase portrait. In the coordinates used for the numerical
transport diagnostic, this center is taken to be
\[
(x,y)=\left(\sqrt{p_0/K},0\right),
\]
with radius $R=0.8$. These points are initialized at $t=0$ and integrated forward through the full
nonautonomous equations of motion, Equations~\eqref{eq:x_dot} and~\eqref{eq:y_dot}, to a final
time of $t=8$. Trajectories are explicitly classified by their destiny: ``safe'' if they remain in the
right half-plane $(x(8)>0)$ and ``leaked'' if they cross the dividing surface into the left half-plane
$(x(8)<0)$.

\begin{figure}[!htbp]
    \centering
    \includegraphics[width=0.90\textwidth]{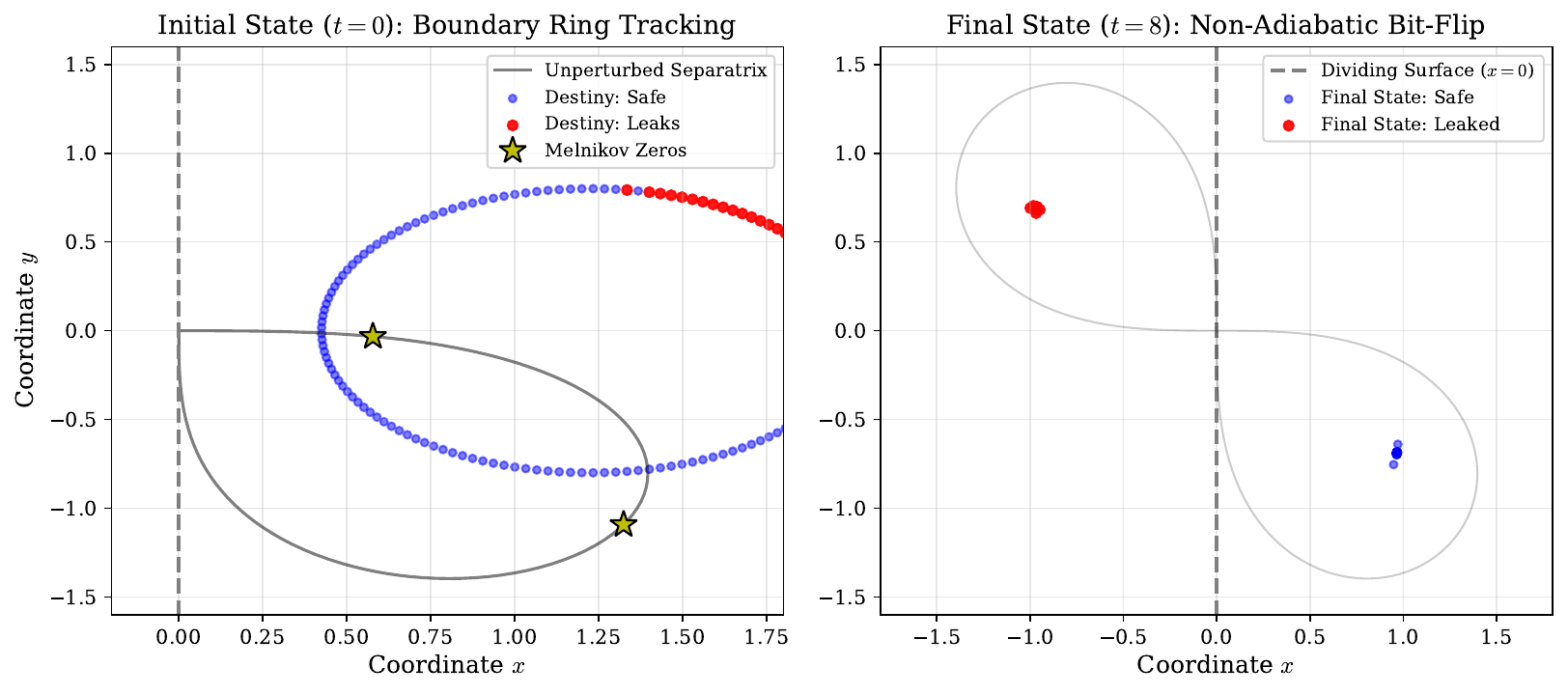}
    \caption{Illustrative full-system simulation of lobe-mediated transport using the full dissipative
    nonautonomous system for a pulse exceeding the Melnikov threshold. (Left) Initial conditions at
    $t=0$, color-coded by their final-state classification at $t=8$. The initial states that eventually
    cross the dividing surface (red) form the turnstile lobe, with yellow stars marking the intersection
    locations inferred from the zeros of the Melnikov function. (Right) Final state at the conclusion of
    the gate pulse, showing that the red cluster has been transported across the $x=0$ dividing surface
    into the opposite logical region.}
    \label{fig:appendix_leakage}
\end{figure}

As demonstrated in Figure~\ref{fig:appendix_leakage} (Left), the initial conditions partitioned by
their final-state classification explicitly trace out the transient homoclinic tangle. The subset of
initial conditions that undergo non-adiabatic leakage (red) form a distinct geometric lobe. Crucially,
the base of this numerically revealed lobe aligns closely with the intersection points inferred
analytically from the Melnikov zeros (yellow stars). Figure~\ref{fig:appendix_leakage} (Right)
confirms that this specific packet of phase-space area is actively transported across the $x=0$
dividing surface by the conclusion of the pulse, corresponding in the semiclassical phase-space picture
to a logical bit-flip event. This connects the analytical Melnikov predictions to the corresponding semiclassical
non-adiabatic transport event.

\FloatBarrier

\end{document}